# Reducing sheet resistance of self-assembled transparent graphene films by defect patching and doping with UV/ozone treatment

**Tijana Tomašević-Ilić[1*], Đorđe Jovanović[1], Igor Popov[1,2], Rajveer Fandan[3], Jorge Pedrós[3], Marko Spasenović[1], Radoš Gajić[1]**

[1]Graphene Laboratory (GLAB) of the Center for Solid State Physics and New Materials, Institute of Physics, University of Belgrade, Pregrevica 118, 11080 Belgrade, Serbia

[2]Institute for Multidisciplinary Research, University of Belgrade, Kneza Višeslava 1, 11030 Belgrade, Serbia

[3]Departamento de Ingeniería Electrónica and Instituto de Sistemas Optoelectrónicos y Microtecnología, Universidad Politécnica de Madrid, Madrid 28040, Spain

*E-mail: ttijana@ipb.ac.rs

**Highlights**

- Photochemical oxidation (UV/ozone treatment) of self-assembled solution-processed graphene films is demonstrated for the first time

- The effect that photochemical oxidation has on self-assembled graphene films is a reduction of sheet resistance, the opposite of the earlier reported effect on CVD or mechanically exfoliated graphene

- Edges as the dominant defect type in self-assembled graphene films play a crucial role in the presented defect density decrease

- Photochemical oxidation is proposed as a method of increasing the carrier mean free path, doping, and lowering sheet resistance of solution-processed graphene films

**Abstract**

Liquid phase exfoliation followed by Langmuir-Blodgett self-assembly (LBSA) is a promising method for scalable production of thin graphene films for transparent conductor applications. However, monolayer assembly into thin films often induces a high density of defects, resulting in a large sheet resistance that hinders practical use. We introduce UV/ozone as a novel photochemical treatment that reduces sheet resistance of LBSA graphene threefold, while preserving the high optical transparency. The effect of such treatment on our films is opposite to the effect it has on mechanically exfoliated or CVD films, where UV/ozone creates additional defects in the graphene plane, increasing sheet resistance. Raman scattering shows that exposure to UV/ozone reduces the defect density in LBSA graphene, where edges are the dominant defect type. FTIR spectroscopy indicates binding of oxygen to the graphene lattice during exposure to ozone. In addition, work function measurements reveal that the treatment dopes the LBSA film, making it more conductive. Such defect patching paired with doping leads to an accessible way of improving the transparent conductor performance of LBSA graphene, making solution-processed thin films a candidate for industrial use.

# 1. Introduction

Graphene, with its high optical transparency and low sheet resistance, is an excellent choice for transparent electrodes in various optoelectronic devices [1]. For such applications, transparency in the visible part of the spectrum should be above 80%, while the sheet resistance should be low enough for practical use, all while keeping production costs to a minimum. In the past decade, numerous research efforts were performed to achieve production of thin graphene films usable in practical applications [2-4]. Although chemical vapor deposition (CVD) yields graphene sheets of high quality that can be scaled for industrial use [5], the method is generally regarded as costly [6] and alternative methods are being sought that satisfy the quality/cost tradeoff. Liquid phase exfoliation (LPE) [7] is the most perspective way of obtaining large quantities of exfoliated graphite in solution at reasonable production costs. Nevertheless, all solution-processed graphene needs to be controllably assembled into thin films of satisfactory quality for transparent conductor applications. A number of film assembly strategies exist, such as evaporation-based assembly, assisted, and micropatterned assembly [8]. Each specific thickness and arrangement of graphene sheets in a thin film directly affects physical properties of the film [9] and device performance. Langmuir-Blodgett (LB) and Langmuir-Schaefer (LS) deposition, based on surface-tension induced self-assembly of nanoplatelets at an interface of two liquids or a gas and a liquid, are prime candidate methods for production of large-scale, highly transparent thin graphene films [10-11]. However, all self-assembled films suffer from a large density of defects that often leads to a high sheet resistance of deposited film. Conversely, the large defect density offers an opportunity for surface treatment such as annealing, chemical doping and functionalization [3, 12], all of which can reduce sheet resistance or produce other desirable effects. The susceptibility of a film to treatment as well as its initial sheet resistance depend on the nature of the prevalent defects, such as impurities, vacancies, nanoplatelet edges, and topological defects, as well as the defect density. For example atoms located at the edges of a graphene sheet exhibit higher reactivity compared to those in the basal plane, making the ratio of the density of edge atoms to basal-plane atoms the determining factor for the efficiency of surface modification [13]. It is thus imperative to carefully study the nature and density of defects in any thin film transparent conductor, especially when considering physical or chemical treatment to enhance the film's practical usability.

Here, we report characterization of the defect type of Langmuir-Blodgett self-assembled (LBSA) films from LPE graphene and subsequent defect patching with UV/ozone (UVO) treatment. We observe the effects that photochemical oxidation has on our films exposed to ozone, a very important gas adsorbate that significantly alters the properties of materials through doping, affecting the performance of electronic devices [14-16]. As shown earlier, oxidation spreads from edges inwards across the entire surface of graphene flakes [17]. When applied even for a short time to mechanically exfoliated and monolayer CVD graphene, UVO leads to significant defect generation resulting in an increase of sheet resistance [18, 19]. Ozone reacts with the edge sites of CVD graphene until reaching a saturation point. Beyond saturation, the basal plane becomes more susceptible to oxidation, resulting in the replacement and relief of carbon atom defects. We find that nanoplatelet edges are the dominant defect type in our films, in contrast to CVD-grown graphene and earlier reported mechanically exfoliated graphene, where charged impurities and covalently bonded adatoms are the limiting factor for carrier mobility [20]. We treat the film surface with UVO and find that the sheet resistance decreases by a factor of 3, while optical transparency throughout the visible part of the spectrum remains high (>80%) and virtually unchanged. Measurements of the surface work function indicate that doping is responsible for the decrease in sheet resistance. FTIR spectroscopy confirms formation of oxygen-containing groups after UVO treatment. With a careful analysis of Raman spectra, we find that the density of defects decreases with treatment, yielding an increase in the carrier mean free path, while edges remain the dominant defect type, all indicating that the ozone binds predominantly to the edges of graphene nanoplatelets. We perform the same UVO treatment on CVD graphene and show that on monolayer CVD graphene, UVO has a detrimental effect on sheet resistance. We also treat thick CVD-grown multilayer graphene films, which prove to be robust against UVO treatment, although such films have very low optical transparency. Furthermore, in order to understand the experimental results we theoretically analyze deposition of an ozone molecule on the edges of a wide graphene nanoribbon (GNR) as a nanosystem that well approximates LBSA film. After we determine the deposition mechanism, we present electronic and transport properties of such oxidized ribbons. Hence, transport and work function measurements indicate increased film doping, AFM indicates that no major macroscopically observable morphological changes are made on the film, Raman resolutely points to edge patching as the dominant interaction mechanism, while FTIR shows that oxygen binding to the graphene lattice occurs during treatment. Our experimental study is firmly backed by ab-initio calculations that indicate that ozone species binding to edges will increase film conductivity. We thus conclude that UVO treatment is a good option for reducing sheet resistance of LBSA LPE graphene films, bringing the electronic performance of these sheets closer to that of CVD graphene which is produced at a higher cost.

## 2. Methods

A graphene dispersion was prepared from graphite powder (Sigma Aldrich, product no. 332461) from a concentration of 18 mg ml-1 in N-Methyl-2-pyrrolidone (NMP, Sigma Aldrich, product no. 328634), exposed to 14 hours of sonication in a low-power sonic bath. The resulting dispersion was centrifuged for 60 min at 3000 rpm in order to reduce the concentration of unexfoliated graphite. The resulting dispersions were used to form films approximately 3 nm thick by LBSA, in the same way that we demonstrated in our previous work [12, 21]. A small amount of graphene dispersion is added to a water-air interface and after the film is formed, it is slowly scooped onto the target substrate (Figure 1a). Glass and SiO2/Si are used as substrates.

For single-layer CVD studies, we used commercially available monolayer CVD graphene grown on 20 μm thick copper foil (Graphene Supermarket) and transferred onto $SiO_2$/Si substrate with a home-built automatic transfer system using ammonium persulfate (($NH_4$)$_2S_2O_8$) 0.3M as copper etchant [22]. For multilayer CVD studies, we used multilayer graphene with an average thickness of 105 nm (about 300 monolayers) grown on 25 μm-thick nickel foil (Graphene Supermarket). We etched away the nickel foil in a 0.25 M solution of ferric chloride ($FeCl_3$) in water, yielding a floating multilayer graphene film which was scooped out of the solution onto a $SiO_2$/Si substrate in the same way as already reported for multilayer graphene condenser microphones [23].

Photochemical oxidation (UVO treatment) is performed by exposing the graphene films to ultraviolet radiation and ozone for 3, 5, 15 and 30 min at a $50^0$C chamber temperature and ambient pressure in a Novascan UV/ozone Cleaner by converting oxygen from ambient air to ozone using a high intensity mercury lamp (Figure1b). We perform the treatment in a standard commercially available UVO cleaner and acknowledge that while varying the intensity of the radiation and/or the concentration of ozone gas may lead to interesting results, it will be part of a subsequent study.

For optical characterization, UV-VIS spectra were taken using a Perkin-Elmer Lambda 4B UV/VIS Spectrophotometer. The oxidation process was characterized using a TriVista 557 S&I GmbH Micro Raman spectrometer ($\lambda$ = 532 nm) at room temperature. FTIR spectra were measured with a Thermo Scientific Nicolet 6700 FT-IR spectrometer in the diffuse reflectance infrared Fourier transform (DRIFT) mode. The resistance of each sample was measured in a two-point probe configuration and the sheet resistance was obtained by considering sample geometry factors.

The work function measurement is performed with Kelvin probe force microscopy (KPFM, NTEGRA Spectra), prior to and after photochemical treatment of our graphene films.

The calculations are done using density functional-based tight binding method (DFTB) [24, 25] with self-consistent charge correction as implemented in the DFTB+ code [26]. Spin polarization was included in calculations. This method has a proven record of various applications to graphene and graphene nanoribbons [27-30]. Transport properties are calculated by DFTB augmented with the Green's functions formalism [31]. Since the atomic structure of LBSA graphene is dominated by edges, we consider GNR a suitable nanosystem that well approximates the nanoflakes in our experiment. For this purpose we model a wide GNR with width 2 nm. The interaction between electronic clouds of two GNR edges is small for such a wide ribbon, hence its electronic properties are equal to the asymptotic limit of wide ribbons [32]. The periodic (infinite) edges of GNR correspond to the flakes in the experiment, which have large circumferences, i.e. long edges. Utilization of GNR instead of nanoflakes per se is not only physically equivalent but also numerically much more tractable.

## 3. Results and discussion

Figure 2a depicts the sheet resistance of graphene films upon exposure to UVO. Prior to exposure, the sheet resistance of LBSA graphene (red circles) is above 80 kΩ/sq. Upon exposure, the sheet resistance decreases rapidly within the first 5 min, reaching a value below 30 kΩ/sq, an approximately 3-fold reduction. At those exposure levels, oxidation reaches a saturation point and remains stable for longer exposure times. Single layer CVD graphene exhibits a pronouncedly different behavior (violet circles), starting from a very low value of sheet resistance which gently rises after 5 mins of exposure, dramatically increasing for longer exposures. After 15 minutes of treatment, sheet resistance of CVD graphene is still 3-4 times smaller than that of LBSA graphene, whereas after 30 minutes of exposure LBSA graphene exhibits 4 times smaller sheet resistance. Multilayer CVD graphene (blue circles) has a sheet resistance of ~ 8 kΩ/sq, which changes only slightly even for long exposures to UVO.

It is expected that CVD graphene compared to LBSA graphene boasts a lower sheet resistance, which is inherently related to carrier mobility. Carrier mobility is inversely proportional to the density of scattering defects, which should be small in CVD graphene. LBSA graphene morphology has an abundancy of nanoplatelet edges [21] that act as scattering centers and have a detrimental effect on initial sheet resistance. However, with UVO treatment, that resistance decreases, pointing to a reaction of ozone with existing defects [17]. On CVD graphene, the few edge sites are quickly fully saturated by ozone molecules, forcing the molecules to deposit their energy by formation of defects or through adsorption, thus creating new point defects on the basal plane, having a detrimental effect on sheet resistance [19]. In multilayer graphene, UVO has little effect on sheet resistance because the ozone molecules react with the top layers only, whereas charge transport takes place through the entire volume of the sheet.

To examine the origins of decreasing sheet resistance with UVO treatment, we measure the surface work function (WF) with KPFM. Figure 2b depicts KPFM maps of the contact potential difference (CPD) between the sample and the tip, before and after UVO exposure. Note the different colormap scale, that indicates a lower CPD between the sample and the metallized AFM tip for treated samples compared to untreated samples. To measure the absolute value of the WF, we use as a reference the work function of highly ordered pyrolytic graphite (HOPG), a tabulated value of 4.6 eV [33, 34]. In order to determine the average CPD of the measured surface, histograms of KPFM maps were used and fitted to a Gaussian distribution (Figure 2c). The mean

WFs of the tip, HOPG, untreated, and treated films are plotted against the vacuum level in Figure 2d. The exact procedure is detailed in our previous work [21]. It is evident that UVO increases the WF from 4.8 eV to 4.9 eV, shifting the Fermi level downwards by ~100 meV. It is this additional UVO-induced p-doping that leads to an increased carrier concentration, in turn causing decreased sheet resistance [35].

Aside from improving the sheet resistance, for transparent conductor applications it is important that the optical transparency of the treated film remains high. Figure 3a depicts AFM topography of the same areas of the film that were used to measure the WF. No difference is observed in the film morphology as an effect of UVO exposure. Figure 3b depicts the transmittance spectrum of the pristine and treated UVO film. The film has good transparency throughout the visible part of the spectrum, with 80-85% transmittance. The transmittance of the film is barely affected by the treatment. For comparison, monolayer CVD would have transmittance on the order of ~97%, whereas multilayer CVD transmits under 10% of light in the visible, according to manufacturer specifications. Hence, single layer CVD graphene has the overall best performance for transparent conductor applications, although cost remains a limiting factor for wide-scale use, whereas multilayer CVD has little use for such applications due to its low transparency. LBSA graphene holds the middle ground, with acceptable electronic and optical performance (especially after UVO treatment) with projected low costs of fabrication and good scalability. It is important to note that the unchanged transparency coupled with a reduced sheet resistance leads to a threefold increase in the figure of merit (FOM) for transparent conductor applications [36].

Figure 4 shows FTIR spectra for untreated and UVO treated films. It is clear that the intensities of peaks associated with oxygen-containing groups increase after UVO exposure. We distinguish a change in shape and total peak area of a broad band in the 3000-3700 cm$^{-1}$ region after UVO treatment, corresponding to the presence of hydroxyl and carboxyl groups as well as water. Also, a peak appears at 1825 cm$^{-1}$ after 30 min of UVO exposure indicating the formation of interacting oxygen groups (OH and C=O). Another evident change after 30 min of exposure occurs near 800 cm$^{-1}$, in a spectral region associated to the formation of epoxides (C-O-C) at graphene edges, according to a previous report [37]. The overlapping spectral features of ethers, epoxides, carboxyles and hydroxyl groups complicate spectral interpretation in the 1000-1300 cm$^{-1}$ region where there is a strong SiO$_2$ absorption band.

The opposite effect that UVO has on CVD and LBSA graphene demands an inquiry into the effect of reactive site morphology on reactions with ozone. LBSA graphene morphology is dominated by nanoplatelet edge defects, whereas CVD yields graphene that has a few edges, in which chemical reactions should be governed by point defects such as charged impurities and covalently bonded adatoms [36]. To clarify the nature of reactive defects and their evolution during UVO treatment, we apply Raman spectroscopy, a versatile tool for characterization of graphene-based materials [38].

Figure 5a depicts Raman spectra of LBSA graphene as a function of UVO exposure. The spectra feature four pronounced bands: D at ~1343 cm$^{-1}$, G at ~1579 cm$^{-1}$, D' at ~1614 cm$^{-1}$ and the 2D band at ~2694 cm$^{-1}$. Furthermore, several combinations of these bands are also observed: D + D" at ~2450 cm$^{-1}$ and D +D' at ~2935 cm$^{-1}$ (Figure 5b), where D" is signature of a phonon belonging to the LA branch, typically observed at ~1100 cm$^{-1}$ [38]. The D and G bands are well resolved for all samples. The 2D peak is a typical signal arising in multilayer graphene. However, Raman spectra show evident changes of the intensity of the D mode with UVO exposure. Figure 5c depicts the ratio of the D peak to the G peak calculated from integrated peak areas, often used to monitor defect evolution in graphene. We observe a large decrease of this ratio during UVO exposure, indicating a reduction in defect density. The D/G intensity ratio evolution shows the same trend as sheet resistance, with a rapid change within the first 5 minutes of exposure, followed by saturation. The reduction of defect density with UVO exposure in LBSA graphene is opposite from our results on CVD graphene (Figures S1 and S2) and the literature on monolayer graphene [12-13]. LBSA graphene thus responds in a unique way to an oxidizing environment, with ozone binding to existing defects leading to improved electrical performance.

The ratio $I_D/I_G$ can be converted to the carrier mean free path ($L_D$), as long as the laser wavelength is known [39]:

$$L_D^2 (nm)^2 = (1.8 \pm 0.5) \times 10^{-9} \lambda_L^4 \left(\frac{I_D}{I_G}\right)^{-1} \tag{1}$$

For the wavelength used in this study ($\lambda_L$=532nm), we plot $L_D$ as a function of exposure time in Figure 5d. $L_D$ in LBSA graphene increases from 15 nm to 19.5 nm upon UVO exposure, again indicating defect patching. Before and after exposure, the mean free path is smaller than our average flake diameter (previously reported as 120 nm [21]), which points to defects within the nanoplatelets, either through edges of sheets that are stacked on top of each other, or through other point-like defects that we cannot observe with AFM and SEM.

Possible defects in graphene include topological defects (such as pentagon-heptagon pairs), boundaries, vacancies, substitutional impurities, and $sp^3$ defects [40]. Topological defects have the lowest formation energy [41], and they are always present in LPE graphene sheets as a result of the cavitation process [42]. As the ratio between the intensity of the D and the D' mode is very sensitive to the type of defect, with a value of 3.5 for edges, 7 for vacancies, between 7 and 13 for substitutional impurities, and 13 for $sp^3$ defects [43, 44], we measure this value to deduce the nature of defects in our sample. We observe that the ratio of the D-peak intensity to the D'-peak intensity in our films is nearly constant at a value of 4.8±0.5, regardless of UVO exposure. The measured ratio indicates that edges are the dominant defect type in our films, ruling out vacancies, substitutional impurities, and $sp^3$ defects, in agreement with previously published data for LPE graphene [42]. There is little change in defect type with photochemical oxidation (Figure 5e), although the defect density decreases, indicating that ozone most likely reacts with the existing defects and patches them.

In LBSA graphene, flakes may bundle in stacks with varying thickness and lateral dimensions, edge geometries with varying saturation levels (bound oxygen, hydrogen, or other chemical groups), and a wide variety of possible defects. From these virtually infinite possibilities, for our DFTB+ calculation we choose an example of a GNR with bare zig-zag edges and with width of 2 nm. Electronic properties of graphene depend sensitively on physical and chemical modification of edges [45-47]. It was shown that roughness at zig-zag edges does not significantly influence their conductance in contrast to armchair edges [48-50]. The simple choice of a zig-zag GNR is for demonstration purpose only, i.e. to uncover the basic physics of the experimental results, without intent to cover all aspects of the experimental reality. The initial atomic structure consists of the GNR with an ozone molecule placed parallel to the edge. After geometry optimization of the system we obtained a transition configuration (TC) using the dimer method [51]. Optimization of the TC geometry led to the next (meta)stable geometry. The nudged elastic band (NEB) method [52, 53] is used for evaluation of the potential barrier between the configurations. Using this procedure we obtained three (meta)stable configurations. The reaction is presented in Figures 5a and 5b. Firstly, one of the two O-O bonds in $O_3$ breaks, which is followed by a rotation of the O-O dimer around a C-C axis by 180°, as illustrated with the insets in Figure 6a. The product of the reaction is the GNR edge with three separate O adatoms at the nearest sites of the edge (the configuration will be called 3O in the remaining text). The potential barrier for the reaction is $E_f = 1.11$ eV, while the energy for the reverse reaction is $E_r = 4.58$ eV. At a temperature of 50°C, which was used in our experiment, both forward and reverse reactions are impossible. However, the exposure of the sample to UV radiation may assist the reaction. The forward reaction is exothermic with $\Delta E = -3.47$ eV. Two of the three O atoms attached at the edge can associate into an $O_2$ molecule, as indicated in Figure 6b. The potential barrier is much smaller in this case, being only $E_f = 0.05$ eV. The reverse reaction is also viable with $E_r = 0.22$ eV. The reaction is exothermic as well, with $\Delta E = -0.17$ eV. The final product of the cascaded reaction consisting of an $O_2$ molecule and an adatom at the GNR edge will be designated as $O_2+O$ configuration in following text. Therefore both configurations, $O_2+O$ and 3O, being nearly isoenergetic, can coexist at the same time. The reaction is somewhat different from the one obtained in [54], but it is expected since the paper considered zig-zag edges in a small hydrogenated carbon cluster in contrast to the infinite ribbon without edge saturation that we consider.

The conductance of the three configurations (the initial $O_3$ and the two reaction products $O_2+O$ and 3O) is shown in Figure 6c. While a pristine zig-zag GNR (not shown in the figure) does not have an electronic band gap, a gap opens upon deposition of $O_3$ at a GNR edge. We set the Fermi level ($E_F$) to the top of the valence band in the initial $O_3$ and $O_2+O$ configurations. The electronic gaps are around 0.10 eV and 0.35 eV, respectively, whereas the conductance at $E_F$ is one quantum for both cases. This corresponds to a sheet resistance of 12.9 kΩ. Breaking of the O-O bond in the $O_2+O$ configuration causes a significantly different conductance. The 3O configuration does not exhibit a gap, while the conductance is increased by a factor of 3 or 5, since $E_F$ is positioned at the border of regions with two distinct conductance values. The 3- to 5- fold increase of conductance of a free-standing GNR indicates a possible mechanism of decreasing sheet resistance in our experiment. Note that there is not a one-to-one correspondence between experiment and theory as transport through LBSA graphene is more complex, involving stacks of GNRs with a variety of edges, and transport between stacks which is perhaps based on quantum tunneling. However the increase of conductance is in positive correlation with our experiments. All configurations exhibit p-type doping, in agreement with experiment.

A complete analysis of electronic properties of stacks of flakes (or GNRs) would require consideration of flakes with numerous combinations of edge structures, physical dimensions of flakes and their different mutual orientations, varying number of GNRs in stacks. After obtaining the conductance inside and between stacks a set of stacks could be modeled with an equivalent electrical circuit and their resistance calculated. This comprehensive study will be published elsewhere. Here, for demonstration only, we opt for a single structure, a Bernal stacking of four GNRs with an $O_2+O$ edge configuration as shown in Figure 6d. We compare projected density of states (PDOS) of the stack in the plane of the top layer, without and with oxidation in Figures 6e and 6f, respectively. The O adatoms responsible for the in-plane oxidation make epoxy groups as found previously in [55]. Upper four sub-graphs in each panel represent PDOS at the GNR edge, while the lower four sub-graphs show PDOS at the center of GNR. Colors of the graphs distinguish four GNRs in the stack (named layers in Figure 5e and 5f). Both edges and the middle of GNRs contribute to PDOS in the range from $E_F - 0.5$ eV to $E_F + 0.5$ eV. Edges have much larger contributions for energies below $E_F - 0.5$ eV, hence conductance is expected to be dominated by edges for a bias voltage larger than ~1V. The stack without in-plane oxidation exhibits a uniform PDOS among layers, with $E_F$ positioned at the top of the valence band. In contrast, the oxidation of the top layer alternatively dopes the layers with p and n character. Since PDOS at $E_F$ is non-zero in three out of four layers, we expect a much larger low-bias conductance of the in-plane oxidized stack than the one without in-plane oxidation. Note that the transport properties analyzed in this section are applicable only to the plane parallel to GNRs in the stack. The conductance in the perpendicular direction has a tunneling character, with expected much larger electric resistance.

## 4. Conclusion

In contrast to the degrading effects it has on CVD graphene, photochemical oxidation (UV/ozone treatment) of LBSA graphene leads to a decrease in defect density, which together with doping reduces sheet resistance while retaining high optical transparency. We find that edges are the dominant defect type in LBSA graphene, with little influence of other defects such as vacancies, impurities, or $sp^3$ defects that are common limiting factors for electronic performance of CVD and mechanically exfoliated graphene. Our study thus shows that graphene film morphology and defect landscape play a crucial role in the effect that UV/ozone treatment has on the film. This accessible treatment improves the performance of LBSA graphene, which is key to embedding such materials in durable devices especially those involving direct exposure to ultraviolet radiation and ozone gas. Our novel observation is expected to contribute to the technological acceptance of thin films based on solution-processed 2D materials.

**Acknowledgments** This work is supported by the Serbian MPNTR through Projects OI 171005 and III 45018, by Qatar National Research Foundation through Project NPRP 7-665-1-125, and by the Innovation Fund through the Collaborative Grant Scheme. R.F. acknowledges support from European Union's Horizon 2020 Research and Inovation Programme under Marie Sklodowska-Curie Grant Agreement No 642688. J.P. acknowledges support from Spanish MINECO (Grant RyC-2015-18968 and project GRAFAGEN, ENE2013-47904-C3). D. J. And J. P. acknowledge support from the International Erasmus+project 2015-2-ES01-KA107-022648. We acknowledge Davor Lončarević for FTIR measurements.

## List of figures

**Figure 1** (a) Schematic of Langmuir-Blodgett self-assembly (LBSA) on a water-air interface (1: film formation, 2: substrate immersion, 3: film deposition), NMP is N-Methyl-2-pyrrolidone; (b) Schematic of film exposure to photochemical oxidation.

**Figure 2** a) Sheet resistance of LBSA, CVD-grown single- and multilayer graphene films as a function of UVO exposure time, b) KPFM map of HOPG, untreated, and UVO treated LBSA film, (c) KPFM histograms of LBSA graphene film before (grey) and after (red) UVO treatment, (c) Schematic of the relation between measured CPDs and their corresponding work functions.

**Figure 3** (a) 3x3 µm$^2$ topograph of HOPG, untreated, and UVO treated LBSA film, b) Transmittance of a LBSA graphene film in the visible range on a glass substrate, before (black) and after (red) 5 min exposure to UVO.

**Figure 4** Reflection infrared spectra of untreated (black), 5 minutes (blue) and 30 minutes (pink) UVO exposed LBSA graphene films. We indicate vibrational modes for hydroxyls (possible C-OH, COOH and H$_2$O contributions) at 3000-3700 cm$^{-1}$, carbonyl (C=O) and carboxyls (COOH) at 1600-1750 cm$^{-1}$, sp$^2$-hybridized (C=C) at 1580 cm$^{-1}$ and epoxides (C-O-C) at 800 cm$^{-1}$. Vibrational modes of SiO$_2$ (LO and TO) appear at 1250 and 1080 cm$^{-1}$, respectively. The areas marked with rectangles are regions in which the most prominent spectral changes occur after UVO exposure.

**Figure 5** (a) Representative Raman spectra of LBSA graphene as a function of UVO exposure time. We recorded Raman spectra in various regions to eliminate spot-to-spot variations in the obtained spectra. The spectra show four main bands: D, G, D', and 2D, as well as some weak combinations of these modes. (b) Close-up view of the main bands, (c) The D/G intensity ratio of the films as a function of UVO exposure. The D/G ratio is calculated from integrated peak areas. (d) Interdefect distance as a function of UVO exposure time. The error bar represents the standard deviation of five measurements, (e) The D/D' ratio as a function of UVO exposure time.

**Figure 6** (a) and (b) Potential energy surface of the cascaded reaction of an $O_3$ molecule with a GNR zig-zag edge ($O_3$ configuration). Breaking of an O-O bond in the ozone molecule and rotation of O-O dimer about C-O dimer (transition state), which settles the intermediate (3O) state, are shown in (a). The consequent association of two neighboring O atoms, yielding the $O_2+O$ configuration, is presented in (b). The reaction is analyzed only on one edge (the left edge in the insets). (c) Conductance at zero bias of the GNR with the three configurations. (d) Illustration of the stack with four layers of GNRs with $O_2+O$ edge configuration doped in-plane by five oxygen atoms. Projected density of states of the stack without (e) and with (f) in-plane doping. Panels (e) and (f) distinguish contributions to PDOS from GNR edges and lines of central carbon atoms.

**Figures**

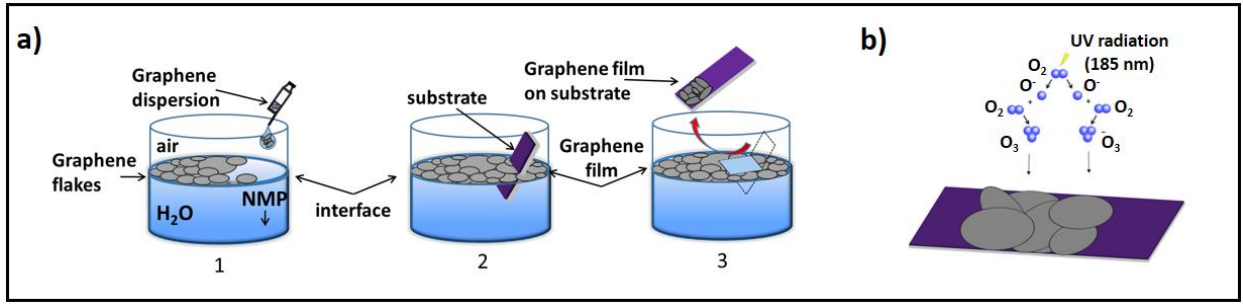

**Figure 1**

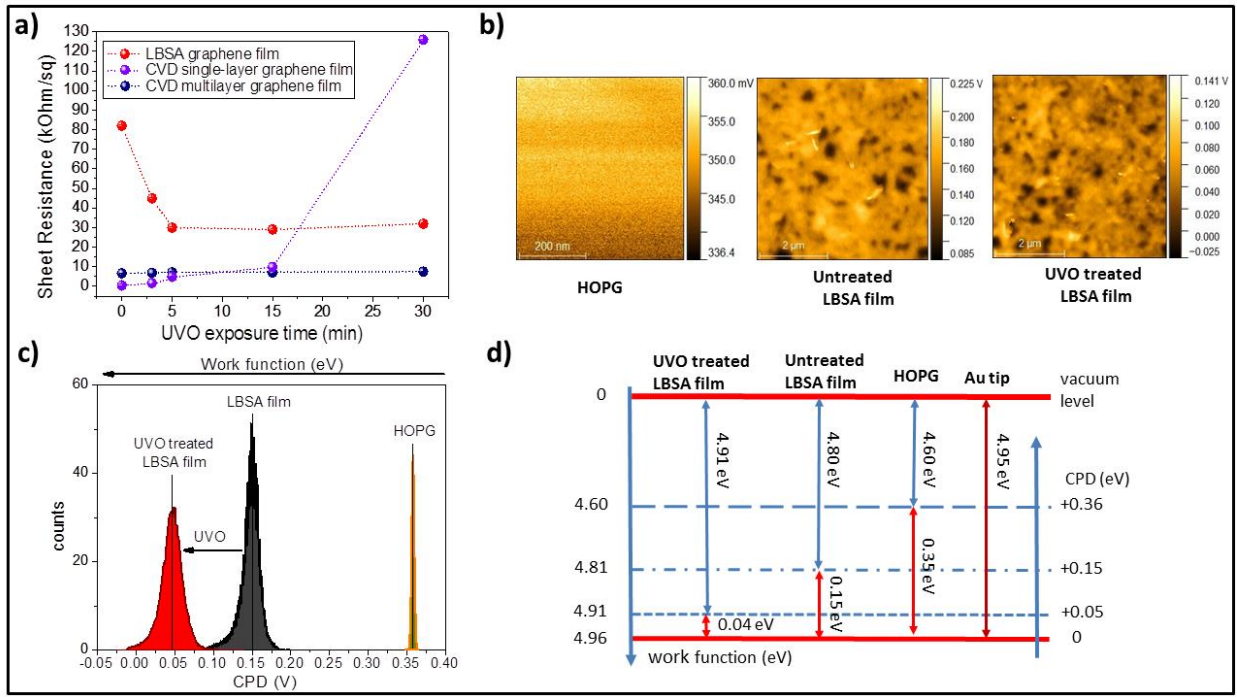

**Figure 2**

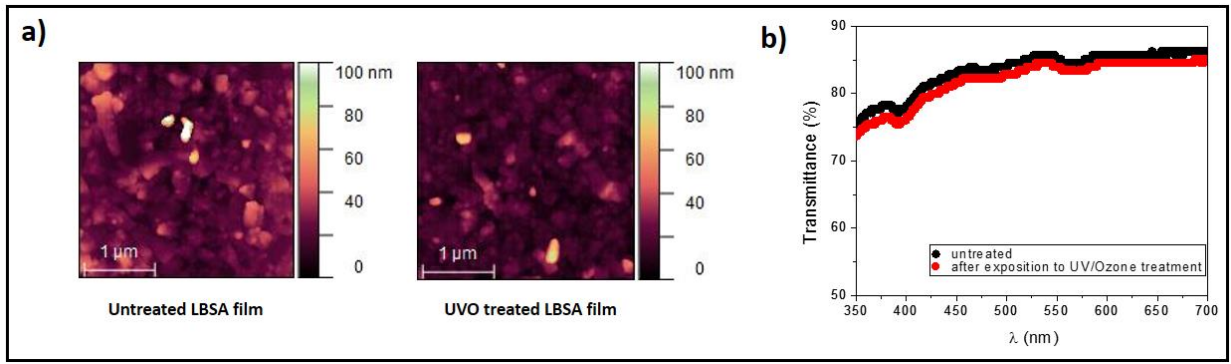

**Figure 3**

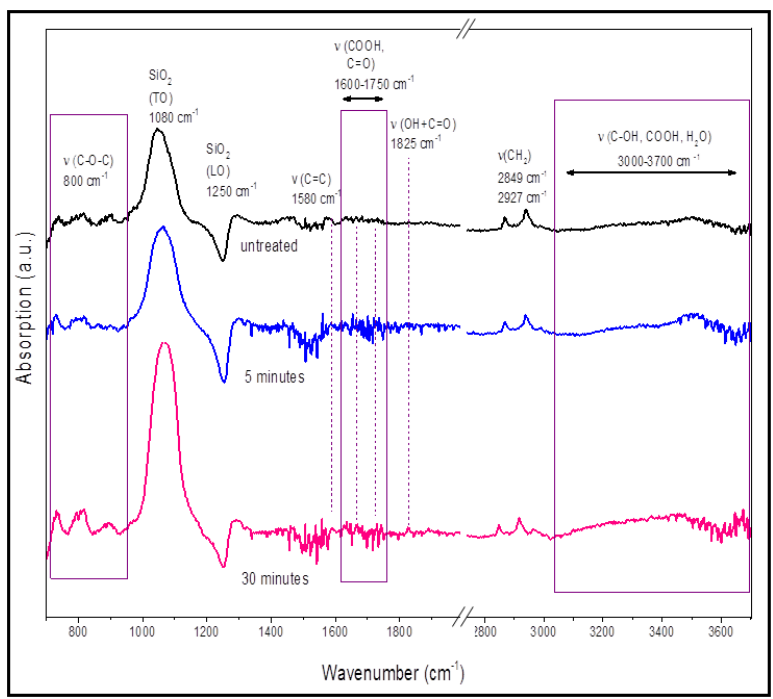

**Figure 4**

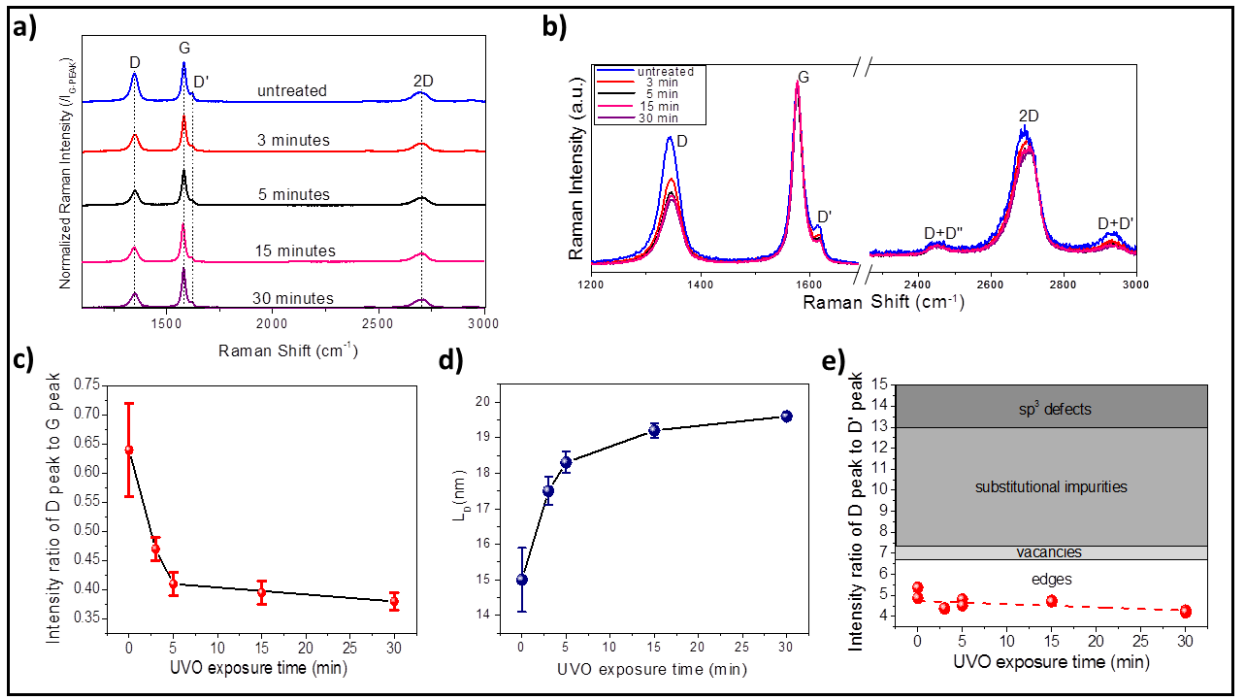

**Figure 5**

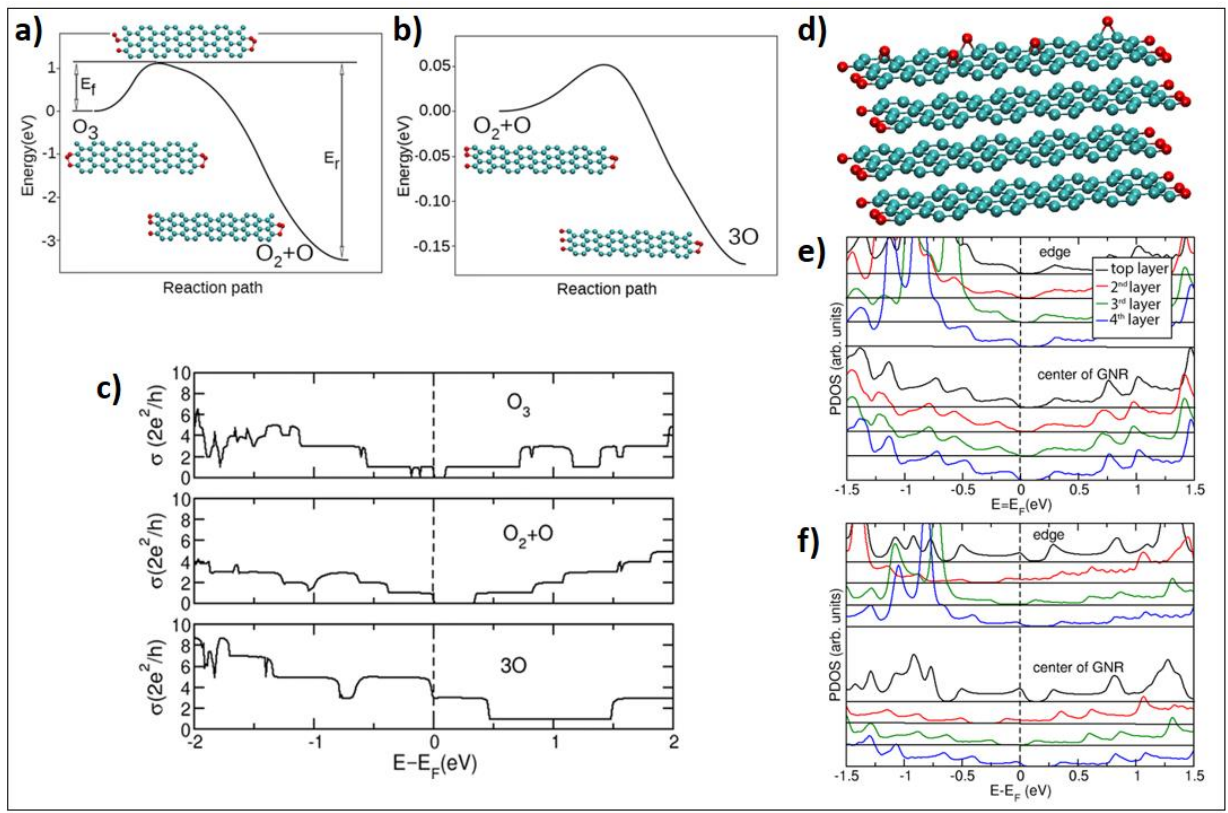

**Figure 6**

# Supplementary information

# Reducing sheet resistance of self-assembled transparent graphene films by defect patching and doping with UV/ozone treatment

Tijana Tomašević-Ilić[1*], Đorđe Jovanović[1], Igor Popov[1, 2], Rajveer Fandan[3], Jorge Pedrós[3], Marko Spasenović[1], Radoš Gajić[1]

[1]Graphene Laboratory (GLAB) of the Center for Solid State Physics and New Materials, Institute of Physics, University of Belgrade, Pregrevica 118, 11080 Belgrade, Serbia

[2]Institute for Multidisciplinary Research, University of Belgrade, Kneza Višeslava 1a, 11000 Belgrade, Serbia

[3]Departamento de Ingeniería Electrónica and Instituto de Sistemas Optoelectrónicos y Microtecnología, Universidad Politécnica de Madrid, Madrid 28040, Spain

*E-mail: ttijana@ipb.ac.rs

UV/Ozone treatment was performed on CVD single- and multilayer graphene. Corresponding Raman spectra are shown in Figure S1.

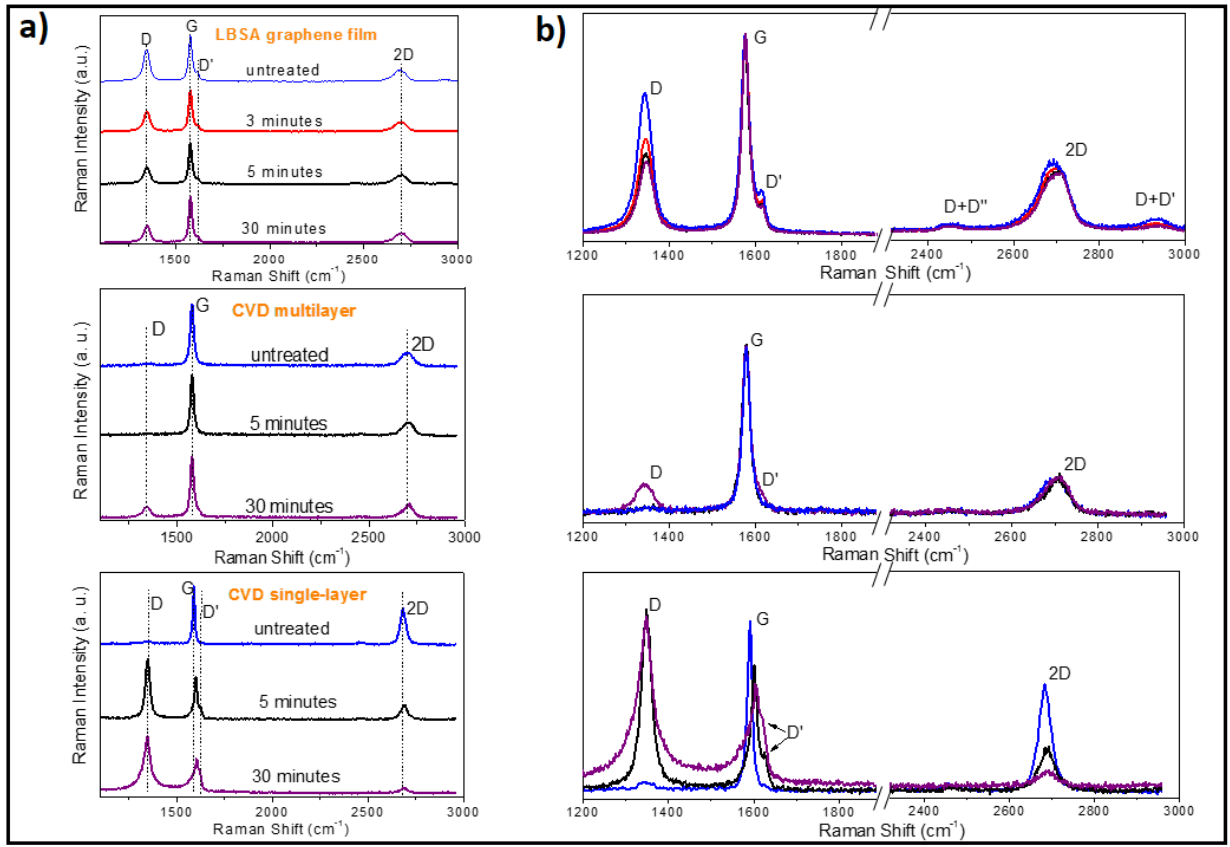

**Fig. S1** (a) Raman spectra of the LBSA, CVD-grown multilayer and CVD-grown single-layer graphene films as a function of UVO exposure time, (b) Close-up view of the main bands of (a).

In all cases, Raman spectra show evident changes of the intensity of the D mode with photochemical oxidation. Figure S1 highlights the different behavior of LBSA and CVD graphene in an oxidizing environment. In contrast to the effect that UVO has on LBSA graphene, described in the main text, treatment of CVD graphene leads to defect generation. Upon photochemical oxidation, CVD graphene develops a strong D peak in the Raman spectra after only 5 minutes of exposure, indicating oxidative removal of the graphitic material and the formation of a significant number of defects. Figure S2a shows a comparison of the ratio of the D-peak to the G-peak before and after UVO exposure for LBSA, CVD multilayer and CVD single-layer graphene. As highlighted in the main text, a large decrease of this ratio is observed after UVO exposure for

LBSA graphene and an increase of this ratio is observed after the same time of exposure for CVD graphene, indicating defect patching in LBSA and defect formation in CVD. Figure S2b shows a change in the mean free path with UVO exposure of all three types of graphene.

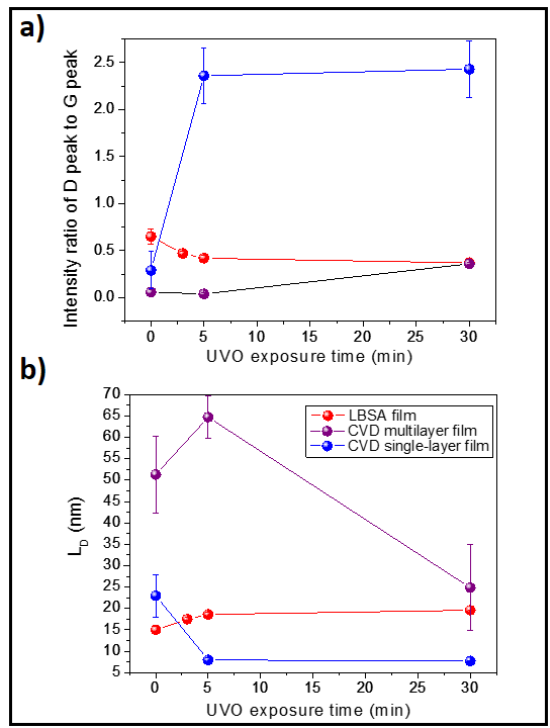

**Fig. S2** (a) The D/G intensity ratio of the LBSA, CVD-grown multilayer and CVD-grown single-layer graphene films as a function of UVO exposure time. The D/G ratio is calculated from integrated peak areas. (b) Interdefect distance of the coresponding graphene films as a function of UVO exposure time. The error bar represents the standard deviation of five measurements.

**ToCG**

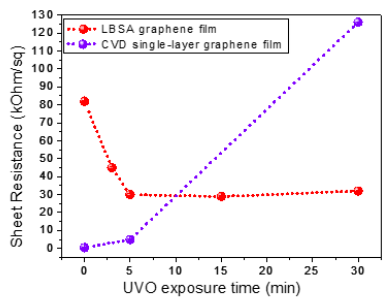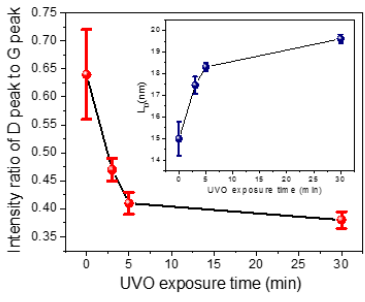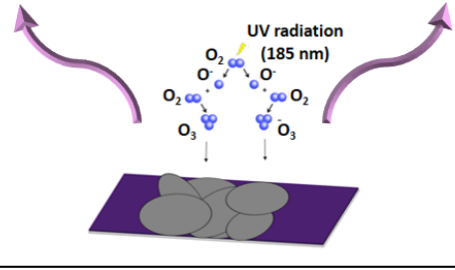